# Fueling the Next Quantum Leap in Cellular Networks: Embracing AI in 5G Evolution towards 6G

Xingqin Lin[†], Mingzhe Chen[§], Henrik Rydén[†], Jaeseong Jeong[†], Heunchul Lee[†],
Mårten Sundberg[†], Roy Timo[†], Hazhir S. Razaghi[†], H. Vincent Poor[§]
[†]Ericsson, [§]Princeton University
Contact: xingqin.lin@ericsson.com

*Abstract*—Cellular networks, such as 5G systems, are becoming increasingly complex for supporting various deployment scenarios and applications. Embracing artificial intelligence (AI) in 5G evolution is critical to managing the complexity and fueling the next quantum leap in 6G cellular networks. In this article, we share our experience and best practices in applying AI in cellular networks. We first present a primer on the state of the art of AI in cellular networks, including basic concepts and recent key advances. Then we discuss 3GPP standardization aspects and share various design rationales influencing standardization. We also present case studies with real network data to showcase how AI can improve network performance and enable network automation.

## I. INTRODUCTION

Since the 3rd Generation Partnership Project (3GPP) completed the first release of the 5th generation (5G) specifications in 2018, we have witnessed a rapid uptake of 5G commercial rollouts worldwide [1]. As we are moving towards 5G-Advanced and the sixth generation (6G) systems, it is anticipated that embracing artificial intelligence (AI) has the potential to help fuel the next quantum leap in cellular networks [2][3].

Cellular networks, such as 5G systems, are becoming increasingly complex for supporting various deployment scenarios and applications [4]. The evolution of 5G systems for improving performance and addressing new use cases makes 5G even more sophisticated. Several technical challenges exist in such complicated 5G systems, from system modeling to algorithm development to network configuration and management. In addition, 5G systems generate humongous data. The use of rapidly advanced AI techniques, such as machine learning (ML) and data analytics, is critical to managing the complexity, identifying patterns in data, optimizing network design, enhancing system performance, and reducing operating costs [5]. The potential of AI techniques, especially deep learning, has led to significant interest from academia in applying them to communication networks (see, e.g., [5][6][7] and references therein). The mobile industry has also been accelerating the use of state-of-the-art AI techniques in cellular networks. However, to a large extent, the current use of AI techniques in real cellular networks primarily relies on proprietary implementations and solutions.

Global standardization has enabled mobile technologies to scale and flourish over the past decades. We anticipate that standardization will again play a key role in accelerating the AI applications to cellular networks. 3GPP has been studying AI-enabled radio access networks (RANs) in Release 17, including the principles, functional framework, use cases, and solutions [8]. However, the scope of the study does not cover air interface design. To reap the full benefits from AI techniques in next-generation RAN, we must optimize the integration of AI into the system design and realization, including the air interface. The 6G standardization is expected to start in 3GPP around 2025. We have a window of opportunity until then to achieve an initial globally aligned way forward on integrating AI into the design of cellular networks. Therefore, despite that the design of the 5G air interface has largely been frozen, there is an urgency to get going and move forward to investigate AI-enabled air interface for 5G evolution in, e.g., 3GPP Release 18, which will start at the beginning of 2022 [10].

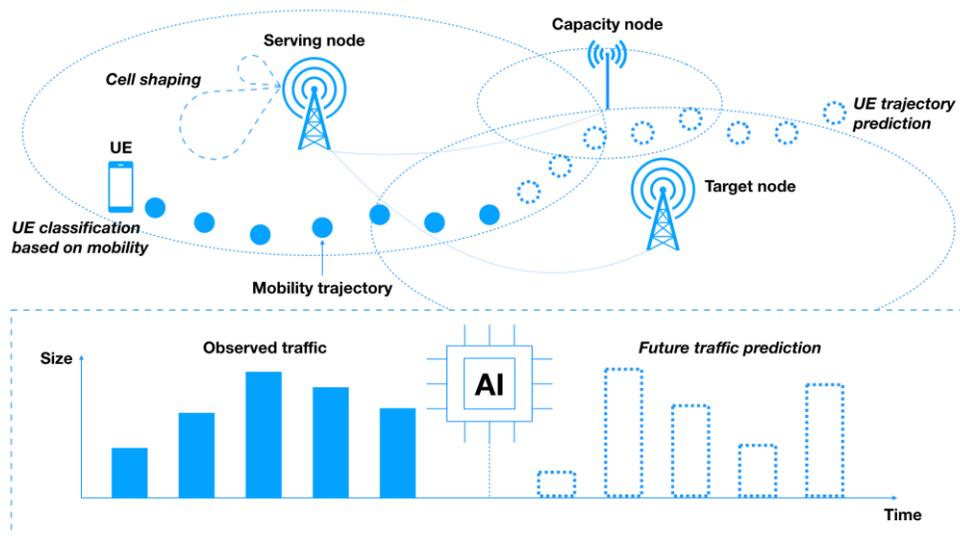

**Figure 1: An illustration of AI use case examples in cellular networks.**

In this article, we first present a primer on the state of the art of AI in cellular networks. Then we discuss 3GPP standardization aspects and share various design rationales influencing standardization. We also present case studies with real network data to showcase how AI can improve network performance and enable network automation.

## II. A PRIMER ON AI IN CELLULAR NETWORKS

### A. AI for Cellular Networks

In cellular networks, AI enables servers, cloud, base stations (BSs), and devices to simulate human intelligence to make predictions, decisions, or identifications. ML is one of the major branches of AI. It enables nodes in cellular networks to perform tasks without being explicitly programmed. Depending on the available data, ML can be generally categorized into three types: supervised learning, unsupervised learning, and reinforcement learning.

Supervised learning requires a labeled dataset in which each data sample consists of an input vector and a corresponding output vector. For example, as illustrated in Figure 1, a BS using a device's historical traffic data to predict the device's future traffic pattern can be considered as a supervised learning task. Unsupervised learning has an unlabeled dataset in which each data sample consists of only an input vector and does not have an output vector. In Figure 1, classifying mobile users according to their mobility pattern can be considered as an unsupervised learning task. Reinforcement learning (RL) enables nodes to learn the control and management strategies, such as resource allocation schemes, by interacting with their dynamic wireless environment. In Figure 1, the cell shaping optimization can be cast as an RL task. One major branch of ML is deep learning (DL), which employs artificial neural networks to analyze high-dimensional data to gain insights and form solutions.

### B. Cellular Networks for AI

The paradigm in which multiple devices jointly perform an ML algorithm stems from the fact that computing capabilities have been built into tens of billions of devices nowadays. These devices can be interconnected, providing platforms for implementing large-scale learning tasks and generating large amounts of data. Training distributed ML algorithms requires the participating devices to exchange their trained ML parameters iteratively. It is essential to jointly optimize network and learning parameters to improve the performance of distributed ML.

Network resource management is an essential aspect of the deployment of distributed ML in cellular networks. One popular approach to finding the optimal resource allocation scheme is first to analyze the convergence bound of distributed ML, figure out how a resource allocation scheme affects the convergence of distributed ML, and then identify the optimal resource allocation scheme. However, since this approach optimizes the convergence bound of distributed ML, it is not guaranteed that an optimal resource allocation scheme can be identified. Another popular method for optimizing resource allocation schemes is applying RL to analyze the training process of distributed ML.

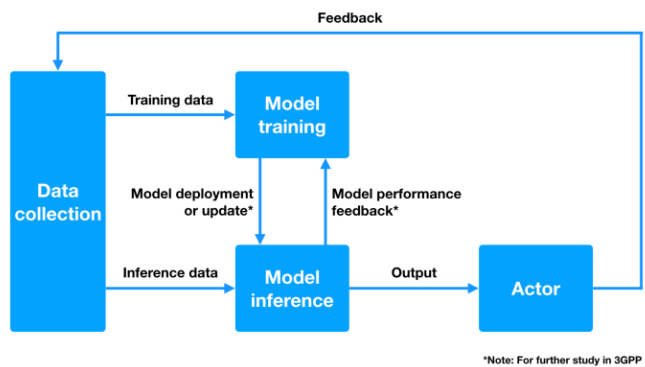

**Figure 2: Functional framework for 3GPP AI-enabled RAN.**

Due to limited energy and computational resource, each device can train and transmit a size-limited ML model. Novel training schemes over cellular networks are needed to support the deployment of distributed ML. One promising training scheme is consensus-based training which enables the devices to implement the distributed training with less reliance on a central controller. The consensus-based training can also be used in a hierarchical distributed ML network, where devices first transmit their trained ML parameters to edge servers and then the edge servers send the aggregated ML parameters to a central controller. Split learning is another commonly used distributed training scheme. It enables each device to train a partial ML model and transmit the output of the trained model to other devices or the BS to complete the rest of the training.

## III. AI-ENABLED 5G: A 3GPP PERSPECTIVE

There have been several 3GPP initiatives on AI-enabled 5G. The initiatives span from RAN to services and system aspects (SA) to core and terminals (CT). Table 1 provides an overview of the AI-related activities in 3GPP. In this section, we focus on the latest 3GPP study item on AI-enabled RAN in Release 17. The objective of the study is to investigate the impact of AI on 5G standardization from a use-case centric perspective. For each identified use case, the study aims to characterize the impact of AI on the node or function in the next-generation RAN (NG-RAN) architecture, e.g., the changes needed in the network interfaces to convey input data to an AI-based function and the signaling of output data from the AI-based function.

### A. Principles of 3GPP AI-enabled RAN

3GPP adopts a set of key principles for AI-enabled RAN to focus on AI applications with potential standardization impact for RAN intelligence. One key principle is that the detailed AI model (e.g., whether to use a decision tree or a neural network model) is not within the scope of the study. Instead, the study should mainly focus on identifying the essential information needed or produced by an AI model for each use case.

Compared to classical algorithmic development, a distinct nature of an AI model is its learning nature. Hence, the 3GPP study investigates the standard impacts of both model training and inference phase. But the details of the training of the AI model, such as the number of training iterations and how the training should prioritize certain data samples, are not within the 3GPP scope.

| Technical Specification Group | Release | Study Item / Work Item | Working Groups | Technical Report |
|---|---|---|---|---|
| **RAN (Radio Access Network)** | Rel-16 | RP-1181456: RAN-centric data collection and utilization for NR.<br>Objective: Investigate RAN-centric data collection and utilization for NR and LTE including use cases and standard impact. | RAN3, RAN2 | TR 37.816 |
| | Rel-17 | RP-201304: Study on further enhancement for data collection.<br>Objective: Study functional framework for RAN intelligence enabled by further enhancement of data collection and identify potential standardization impacts on current NG-RAN nodes and interfaces. | RAN3 | TR 37.817 |
| | Rel-17 | RP-193255: Enhancement of data collection for SON/MDT in NR.<br>Objective: Specify data collection enhancement in NR for self-organizing networks and minimization of drive tests. | RAN3, RAN2 | n/a |
| **SA (Service & System Aspects)** | Rel-16 | SP-180792: Study of enablers for network automation for 5G.<br>SP-181123: New WID on enablers for network automation for 5G.<br>Objective: Study the necessary data to expose to network data analytics function (NWDAF) and the necessary NWDAF outputs; specify system enhancements for NWDAF. | SA2 | TR 23.791 |
| | Rel-17 | SP-200098: Study on Enablers for Network Automation for 5G - phase 2.<br>SP-200975: Work item on enablers for network automation for 5G - phase 2.<br>Objective: Further study NWDAF and specify system enhancements for NWDAF. | SA2 | TR 23.700-91 |
| | Rel-17 | SP-200722: Study on security aspects of enablers for Network Automation (eNA) for 5GS Phase 2.<br>SP-210837: Security aspects of enablers for network automation (eNA) for the 5G system (5GS) Phase 2.<br>Objective: Identify security issues and requirements and provide corresponding security solutions; specify security protection mechanisms. | SA3 | TR 33.866 |
| | Rel-17 | SP-190930: New study on enhancement of management data analytics service.<br>SP-210132: New WID on Enhancements of Management Data Analytics Service.<br>Objective: Study & specify enhancements of management data analytics service. | SA5 | TR 28.809 |
| | Rel-18 | SP-191040: Study on traffic characteristics and performance requirements for AI/ML model transfer in 5GS.<br>SP-210587: New WID on AI/ML model transfer in 5GS (AMMT).<br>Objective: Study use cases, service and performance requirements for 5G system support of AI/ML model distribution and transfer; specify service and performance requirements. | SA1 | TR 33.852 |
| **CT (Core Network & Terminals)** | Rel-16 | CP-192259: CT aspects on enablers for network automation for 5G.<br>Objective: Specify stage 3 procedures on protocol enhancements for NWDAF. | CT3, CT4 | n/a |
| | Rel-17 | CP-211191: Revised WID on enablers for network automation for 5G - phase 2.<br>Objective: Specify the protocols and specifications required to fulfil system enhancements for NWDAF. | CT3, CT4 | n/a |

**Table 1: A summary of 3GPP work on AI-enabled 5G.**

The 3GPP study also outlines a functional framework for the different functionalities in the AI-based RAN application to provide guidance when discussing the use cases. As illustrated in Figure 2, the framework outlines a set of functionalities, including model training, model inference, actor, and data collection. How to communicate between these functions is described on a per use case basis. Although a functional framework is outlined, how the AI functionality resides within the current RAN architecture depends on deployment and specific use cases, which is another key principle of AI-enabled RAN agreed in 3GPP.

### B. Use Cases of 3GPP AI-enabled RAN

The 3GPP study focus can be grouped into two use case areas: Traffic steering and RAN energy efficiency. Both use cases can involve using forecasted traffic and user equipment (UE) trajectories from an AI-based function. The potential standard impact is the signaling support for such AI-based forecasts. The UE trajectory prediction intelligence can reside at the UE or the network. Figure 1 provides an overview of the nodes and predictions involved in the use cases.

**Traffic steering.** This use case has been explored from both a UE's perspective and a RAN's perspective. From a UE's perspective, the target is to find the best serving cell(s) for the UE, often referred to as mobility optimization. The serving node of the UE can use the predictions of the UE location to perform better handover decisions. The forecasted traffic in the target node can be utilized to predict the expected quality of experience (QoE) for the UE. If the mobility optimization is implemented as an AI model, introducing enriched feedback information of the expected UE's QoE performance in the target node can provide additional training data.

From a RAN's perspective, the objective is to balance the load among the network nodes. By introducing signaling that allows the target node to send its forecasted load to the serving node, the serving node can, for example, handover fewer uses to the target node if the load in the target node is expected to be higher. Besides, using the predictions of UE trajectory enables the serving node to, for example, avoid handover fast-moving UEs to a capacity cell on a higher frequency (as illustrated in Figure 1).

**RAN energy efficiency.** This use case is closely related to traffic steering since it also concerns distributing UEs among network nodes. The objective is to reduce energy consumption while maintaining a satisfactory quality of service.

With traffic forecast, RAN energy efficiency can be improved by, e.g., turning off capacity cells, deactivating some antennas, and triggering cells to enter sleep mode. In Figure 1, the target node indicates its expected traffic to the serving node to facilitate the serving node's decision on activating/deactivating the capacity cell. For example, if the target node is expected to have a low traffic load, the serving node can turn off the capacity cell. Otherwise, the capacity cell can be activated, and the target node can handover traffic to the capacity cell. Also, predictions of UE trajectory are valuable inputs when deciding whether to activate certain energy-saving configurations. In particular, the trajectory predictions for UEs traveling along a certain route can be used to activate capacity cells along the route proactively.

## IV. 5G EVOLUTION WITH AN AI-EMPOWERED AIR INTERFACE

This section describes the next step in the evolution of an AI-enabled RAN – using AI in the RAN air interface. Before discussing standardization aspects, we first motivate why AI is being considered for the air interface.

The air interface of a 5G RAN is composed of many interconnected, interdependent modules that need to work seamlessly and reliably with one another. These modules have been chosen mainly for pragmatic reasons – to decompose a large and intractable air interface design problem into smaller, more manageable issues that can be addressed with classical techniques. AI tools are being rapidly integrated into these modules to enhance RANs operating within existing standards. In more future looking work, AI is being applied to jointly optimize multiple classical modules, including the entire transceiver chain [11]. Another area of high potential for AI at the physical layer is model deficiency, i.e., the model used to derive the classical solutions does not capture important aspects of reality. A potential for AI also arises when a model is known, but the optimal solution is too complex for practical implementations. Utilizing AI can help to approximate the true underlying signal characteristics [12].

### A. Standardizing AI-empowered Air Interface

Given the potential benefits of an AI-empowered air interface, there is broad industry support for a study item in 3GPP Release 18 [10]. The purpose of the study item is to ensure the long-term evolution potential of the 5G air interface and, potentially, prepare for an AI-native air interface in 6G. The study item will take a use-case centric approach, starting from the functional framework and principles outlined in the Release 17 RAN study item (see Section III). The Release 18 study item will likely cover evaluation methodologies, key performance indicators (KPIs), and UE and network involvement. Evaluation methodologies and KPIs relate to how 3GPP enables AI-based evaluations and comparisons to current state-of-the-art solutions. Potential topics may include synthetic and real datasets, simulation scenarios, performance-complexity trade-off analysis, and the design of baseline AI solutions. UE and network involvement relates to how data collection, training, and inference might be performed in the network.

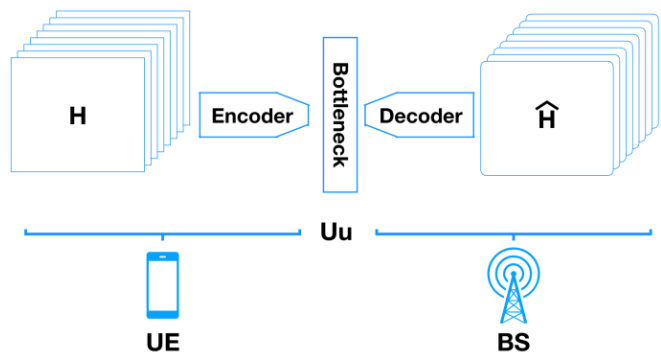

**Figure 3: Autoencoder for CSI compression**

One exemplary use case is channel state information (CSI) reporting, where AI models at both the UE and BS interact with one another over the air interface. In deployments without channel reciprocity, the BS needs the UE to estimate the downlink channel and transmit a CSI report in the uplink. This method of CSI acquisition and reporting has a significant uplink overhead (many parameters are needed to accurately describe the downlink channel). To help reduce uplink overhead, 3GPP has adopted codebook-based CSI feedback solutions for 5G networks. The core idea is that the UE estimates the channel from downlink reference signals and feedbacks, for example, a preferred transmission rank, precoding matrix (from a specified codebook), and a channel quality indicator to the BS. Although efficient, these solutions are suboptimal since many aspects of the downlink channel characteristics are lost in the process.

AI techniques, such as autoencoders, can be used in CSI compression and reporting. The basic principle of an autoencoder is to take an input, perform non-linear compression by an encoder to a lower-dimensional latent representation (sometimes referred to as a bottleneck layer), and decompress the latent representation by a decoder to a defined target. The principle is illustrated in Figure 3 in the context of CSI compression. Here, the UE's estimated channel state ($\mathbf{H}$) is taken as the input to the encoder. The output of the UE's encoder (i.e., the latent representation of the channel in a quantized form) is transmitted over the air interface (Uu) to the BS. The BS uses a matching decoder to reconstruct the channel state ($\hat{\mathbf{H}}$). Hence, the autoencoder model is divided between the UE and BS in this use case, creating an inter-node dependency for model training, model management, etc.

### B. Challenges of Standardizing AI-empowered Air Interface

There are multiple challenges facing the standardization process of an AI-empowered air interface in 3GPP. To concretize, we use the CSI compression and reporting use case to exemplify these challenges.

**Challenge #1: What should 3GPP standardize?**

The autoencoder-based solution for CSI compression and reporting will require a neural network in the UE to interact with a compatible neural network in the BS. The starting point for the 3GPP discussions will likely be that these neural networks are up to implementation (i.e., not standardized). This situation presents several challenges, as described below. The decoder at the BS needs to be trained to match different UE vendor's implementations, but it is not feasible to have a separate model for each UE vendors' implementation. How would such model generalization be realized? How would such a model be trained and maintained? Would it be sufficient to use synthetic data (e.g., 3GPP channel models and scenarios) or data collected in a lab environment? Alternatively, is real data from network operation required? Furthermore, what model configuration and signaling are required to establish a common understanding between UE and BS? These are important issues requiring careful study in 3GPP.

**Challenge #2: How to evaluate AI-based solutions in 3GPP?**

To understand the potential of AI-based solutions, we need to compare them to existing state-of-the-art performance baselines. This poses challenges not encountered in the 3GPP standardization process before. The traditional model-driven signal processing approach usually exhibits inherent robustness to variations in the data observed. The algorithm is typically also explainable, for example, its behavior can be predicted when exposed to different types of data. Both these aspects change when using a data-driven approach. Hence, assuming an AI model for performance evaluation, questions arise relating to how the performance of such model should be evaluated, e.g., what data should the model be trained on? Are the existing synthetic channel models used in 3GPP capturing the richness of the channel sufficiently or is field data needed? In what range of environments should such field data be collected to reflect a rich enough representation? Will field data bias lead to biased solutions where, for example, UEs only work well for particular RAN hardware? What performance baseline should be used for fair performance comparison, and how is complexity compared between an AI-based approach and a more traditional solution? These questions will need to be addressed in 3GPP.

**Challenge #3: How to set performance requirements and conduct conformance testing in 3GPP?**

Existing 3GPP specifications are built on the principle of specifying a set of minimum performance requirements. Such requirements are usually defined in a limited set of conditions, where it is assumed that algorithms will generalize well in real-world deployments. This approach poses a challenge for AI-based algorithms that can be trained and tailored for the set of requirements in the specifications, and hence not generalize. An AI-based solution, which relies on multi-node involvement (such as the autoencoder-based CSI compression and reporting) with each node potentially using a proprietary implementation specific to each vendor, poses further questions. For example, how should one define minimum performance requirements for the equipment under test when the performance depends on specific implementations by other nodes (not available to the testing facility)? Which node is accountable for not fulfilling the performance requirements if multiple nodes contribute to the overall performance? These are important issues requiring careful considerations in 3GPP.

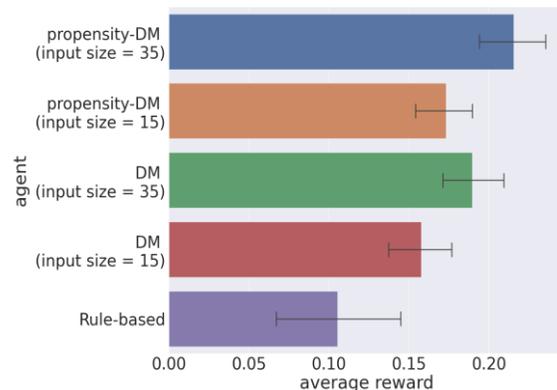

**Figure 4: Performance comparison of offline RL schemes and rule-based policy for antenna tilt optimization.**

## V. CASE STUDIES

AI can be applied to various RAN areas at different time scales, including long-term RAN planning and deployment, short-term RAN management and configuration, and real-time air interface transmission. We presented one specific case study on applying AI to RAN planning and deployment in [13]. In this section, to further demonstrate the potential of AI-enabled RAN, we present two more specific AI applications related to short-term RAN management and configuration and real-time air interface transmission, respectively.

### A. Antenna Tilt Optimization

A RAN typically comprises hundreds of thousands of cells. Antenna tilt can have a significant impact on the coverage and capacity of the network. Optimizing antenna tilts of all cells in the RAN to strike the right balance between coverage and capacity is challenging because it is a large-scale combinatorial problem due to the dependency among the cells.

We take a data-driven approach with RL, where an agent learns a parameter tuning policy from experience. Specifically, we allocate a deep Q-network (DQN) agent to each cell. All agents in the network share the same DQN. Each agent takes as input the KPIs observed at its own and neighbor cells and recommends tilt adjustments for its cell (i.e., uptilt, downtilt, or no change). The selected action (i.e., tilt adjustment) is the one with the highest Q-value among all the actions. The reward of the action is a weighted sum of the KPIs of the agent's own and neighbor cells experienced after the actions are executed.

During the training, a DQN agent explores random actions for learning purpose. However, such exploration is not preferable in a real-world deployment, as the random action trials risk degrading the network performance. Thus, we adopt an offline RL technique, where the DQN agent learns from the history log generated by a baseline policy that has already been deployed and verified in practice. In this way, we can collect training samples without negatively impacting the network performance. We use the samples to train a DQN offline by setting its discount factor to zero, i.e., Q-value is equal to reward, and select the action with the maximum estimated Q-value, which is termed direct method (DM) [14]. But the distribution of the actions in the training samples is biased to the baseline policy, leading to a bias in the trained model using DM. To address the bias, we use a propensity-based learning method to improve DM policy by assigning higher weights to less frequent action samples in each gradient update of the DQN, which we term propensity-DM.

We use a dataset from a real-world RAN [14] that includes observed KPIs of thousands of cells to evaluate the proposed offline RL schemes. In the dataset, tilts of cells were changed daily by a baseline rule-based policy so that the consequence of action could be observed the next day in the log. In Figure 4, we compare the performance of the rule-based approach to the performance of the offline RL schemes, including DM and propensity-DM with different numbers of input features. The results show that the offline RL schemes outperform the rule-based policy, e.g., the propensity-DM policy with 35 input features has a gain of 104% over the baseline rule-based policy. The results in Figure 4 also show that better performance can be achieved with more KPIs as input features.

### B. Beamforming in Multi-cell Multi-antenna Systems

Centralized training and decentralized execution (CTDE) is a promising ML framework for intelligent real-world RAN automation. It allows for non-real-time training based on global information and real-time execution based on local information. To illustrate the concept, we consider the beamforming problem in multi-cell multi-antenna systems. This problem is challenging because of the partial observability at each BS due to practical limitations of accessible information by local agents distributed in the multi-cell systems.

We take a data-driven approach and apply AI to solve beamforming problems. We use a multi-agent framework in which decentralized agents (i.e., actors) with partial observability can learn a multi-dimensional continuous policy with the aid of one shared critic which has access to global information. To fulfill the real-time requirement of beamforming schemes, each actor chooses a beamforming vector based on the local information at its associated BS, where the contextual data on the state of channels is collected. The shared critic learns an action-value function in a centralized manner based on the global information about the channel states and the actors' policies and uses it to learn the action policies that maximize the expected collective reward.

Figure 5 presents a numerical example with two multi-antenna BSs, each serving a single-antenna user. For benchmark purposes, we also plot the rate region, which is the set of all rate pairs that the two users can simultaneously achieve. In this case, we trained two actors and one critic. The critic aims to achieve a Pareto-optimal beamforming strategy by maximizing the collective reward, defined as a weighted sum of the achieved rates of the two users. We denote by α (a value between 0 and 1) the weighting scalar for the rates of the two users. The dashed lines in Figure 5 depict the corresponding learning behaviors under different weighting scalars. We can see a performance gap between the rate pairs achieved by the naïve learning implementation and the Pareto-optimal rate pairs. To close the gap, we propose a feature engineering method, called phase ambiguity elimination (PAE), as a pre-processing step on the input of channel states to improve the learning performance [15]. As shown by the solid lines in Figure 5, the improved implementation with PAE can learn a Pareto-optimal beamforming strategy. The takeaway from addressing the phase ambiguity issue in this case study is that applying AI with deep domain knowledge of the underlying communication technology is crucial for achieving optimal performance in AI-enabled RAN.

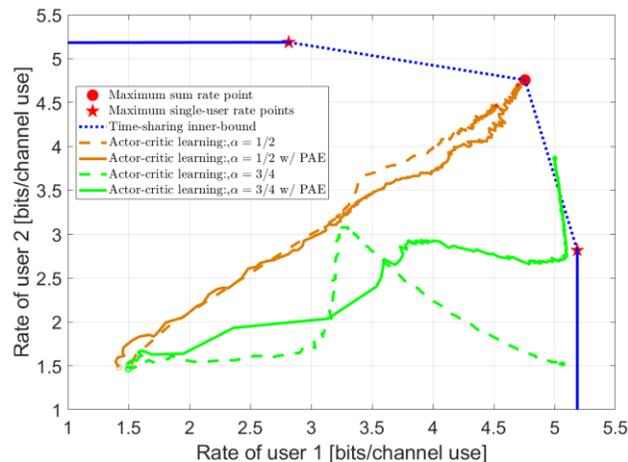

**Figure 5: An illustration of the actor-critic learning behaviors for beamforming in a two-cell multi-antenna system.**

## VI. CONCLUSIONS AND 6G OUTLOOK

The evolution of 5G systems adds ever-increasing complexity and puts numerous demands on cellular networks. AI adoption is essential to efficiently manage the complexity, optimize system performance, and meet the needs of diverse deployment scenarios and use cases. This article has provided a state-of-the-art overview of AI in cellular networks by describing the basic concepts, reviewing the recent key advances, discussing in detail the 3GPP standardization aspects, and sharing various design rationales influencing standardization. This article also sheds light on how AI can improve network performance and enable network automation by presenting case studies with real network data.

As 3GPP is ramping up AI items for the 5G evolution, we anticipate that AI adoption in cellular networks will be accelerated beyond proprietary solutions in the years to come. In particular, the standardization efforts on incorporating AI in the 5G evolution will help achieve an initial globally aligned way forward on integrating AI into the design of cellular networks. The outcome will lay a strong foundation for integrating AI into the 6G system design when the 6G standardization starts in 3GPP around 2025. We will have the opportunity to design the 6G system based on a data-driven architecture that will utilize enormous data volume to support AI across cloud, core, RAN, and device. It will be exciting to see the next quantum leap in cellular networks fueled by AI in the 6G era.

## BIOGRAPHIES

**Xingqin Lin** is a Master Researcher at Ericsson.

**Mingzhe Chen** is a postdoctoral research associate at Princeton University.

**Henrik Rydén** is a Senior Researcher at Ericsson.

**Jaeseong Jeong** is a Senior Specialist at Ericsson.

**Heunchul Lee** is a Senior Researcher at Ericsson.

**Mårten Sundberg** is a Master Researcher at Ericsson.

**Roy Timo** is a Senior Researcher at Ericsson.

**Hazhir S. Razaghi** is a Senior Researcher at Ericsson.

**H. Vincent Poor** is a Professor at Princeton University.